\documentclass[11pt]{article}

\usepackage{amssymb}
\usepackage{amsmath}
\usepackage{amsfonts}

\newcommand{\be}{\begin{equation}}
\newcommand{\ee}{\end{equation}}
\newcommand{\bea}{\begin{eqnarray}}
\newcommand{\eea}{\end{eqnarray}}
\newcommand{\nn}{\nonumber}

\newcommand{\ed}{\end{document}}

\newcommand{\sech}{{\rm sech}}
\newcommand{\eq}[1]{Eq.~(\ref{#1})}
\newcommand{\eqs}[1]{Eqs.~(\ref{#1})}
\newcommand{\e}[1]{(\ref{#1})}
\newcommand{\bt}{\begin{table}\begin{center}\begin{tabular}{|l|l|}\hline}
\newcommand{\et}[2]{\end{tabular} \caption{#1} \label{#2}\end{center}\end{table}}

\begin{document}

\title{Shape Invariant Potentials for Effective Mass Schr{\"o}dinger Equation}
\author{K A Samani\thanks{E-mail address:
samani@cc.iut.ac.ir}~ and F Loran\thanks{E-mail address: loran@cc.iut.ac.ir}  \\ \\
{\it Department of Physics, Isfahan University of Technology (IUT),}\\
{\it Isfahan 84154, Iran}}
\date{ }
\maketitle

\begin{abstract}
Using the method of shape invariant potentials,  a number of exact
solutions  of one dimensional effective  mass Schr{\"o}dinger
equation are obtained. The  solutions with equi-spaced spectrum
are discussed in detail.
\end{abstract}

\section{Introduction}
Since the introduction of  the factorization method by
Shr{\"o}dinger~\cite{1} to solve the hydrogen atom problem
algebraically, there has been a considerable effort to generalize
this method and to find exactly (analytically) solvable potentials
for the Schr{\"o}dinger equation. One of the pioneering works in
this line was done by Infeld and Hull~\cite{factorization}. They
obtained a wide class of exactly solvable potentials using the
factorization method. After the introduction of supersymmetric
quantum mechanics~\cite{witten}, the concept of a shape invariant
potential (SIP) was introduced by Gendenshtein~\cite{sip} in
1983. The Schr{\"o}dinger equation with a SIP has exact solution
and its  energy eigenvalues can be obtained algebraically. The
method of shape invariant potentials is also applied to some
other equations like Fokker - Plank equation as well~\cite{fp}.
Although SIPs are widely studied and many families of them are
obtained~\cite{rev}, but the problem of  the classification of
SIPs has not been completely solved yet.

In this paper we apply the method of SIPs to effective mass
Schr{\"o}dinger equation  and  obtain its exact solutions in a
specific ansatz. Effective mass Schr{\"o}dinger equation was
introduced by BenDaniel and Duke \cite{bd} in 1966 to explain the
behavior of electrons in the junctions of semiconductors. It also
have some applications in heterostructures, graded alloys, and
quantum wells~\cite{ap1}, \cite{ap2}, \cite{ap3}.

Recently, effective mass Schr{\"o}dinger equation is studied by
some  authors and some of its exact solutions  are
obtained~\cite{ex1}, \cite{ex2}, \cite{ex3}. Here we  show that
the method of shape invariant potentials applied to this problem
leads to a wide class of exact solutions.

The paper is organized as follows. In section 2 we review the
method of SIPs  and its generalized form for effective mass
Schr{\"o}dinger equation. In sections 3 and  4 we study two main
classes of generalized shape invariant potentials(GSIPs)
corresponding to special choices for the parameter dependence of
the potential. Such parameter dependence have been  considered to
obtain SIPs for ordinary Schr{\"o}dinger equation in
reference~\cite{rev} and references therein. Due to the importance
of quantum systems with equi-spaced energy levels, we investigate
them separately in each case. Section 5 is devoted to summary and
concluding remarks.
\section{Generalized Shape Invariant Potentials (GSIPs)}
In this section we review the method of SIPs for effective mass
Schr{\"o}dinger equation. Consider an eigenvalue problem given by
    \be
    H\psi_n(x)=E_n \psi_n(x)\;,
    \label{int2}
    \ee
where $H$ is given by
    \be
    H=-\frac{d}{dx}\frac{1}{2m(x)}\frac{d}{dx}+V(x)\;.
    \label{int1}
    \ee
Here $m(x)$ is the position depended effective mass and $V(x)$ is
the potential.

Suppose that the minimum eigenvalue of $H$ is $E_0$. Let us
define a new Hamiltonian $H_1:=~H-E_0$. The Hamiltonian $H_1$ can
be written  in the following form
    \be
    H_1=-\frac{d}{dx}\frac{1}{2m(x)}\frac{d}{dx}+V_1(x)\;,
    \label{f1}
    \ee
in which $V_1(x)=V(x)-E_0$.  The ground state energy of $H_1$ is
zero, i.~e.~ the spectrum of $H_1$ is non negative. Consequently
the Hamiltonian $H_1$ can be assumed to be
    \be
    H_1=A^{\dagger}A\;,
    \label{f2}
    \ee
where $A$ is a first order differential operator of the form
    \be
    A=U(x)\frac{d}{dx}+W(x)\;.
    \label{f3}
    \ee
Inserting  $A$ from \eq{f3} in \eq{f2} and comparing the result
with \eq{f1} one gets
    \bea
    &&\frac{1}{2m(x)}=U^2(x)\;, \label{f3.1}\\
    &&V_1(x)=W^2(x)-\left[U(x)W(x)\right]'\;. \label{f3.2}
    \eea
Here, a  `$'$' indicates the derivative with respect to $x$.
\eq{f3.2} is a generalization of the well known Riccati equation.
We  call the solution of this equation~\cite{rev}, $W(x)$, the
{\it superpotential}.

Next we  define the Hamiltonian $H_2$ as follows
    \be
    H_2:=AA^\dagger \equiv
    -\frac{d}{dx}\frac{1}{2m(x)}\frac{d}{dx}+V_2(x)\;.
    \label{f4}
    \ee
Then the potential $V_2(x)$ reads
    \bea
    V_2(x)&=&W^2(x)+\left[U(x)W(x)\right]'-2U'(x)W(x)-U(x)U^{''}(x)\;,\nn\\
    &&W^2(x)-\left[U(x)W(x)\right]'+2U(x)W'(x)-U(x)U^{''}(x)\;.
    \label{f5}
    \eea
The Hamiltonian  $H_2$ is called the supersymmetric partner of
$H_1$~\cite{rev}. It can be easily shown that $H_1$ and $H_2$ has
the same spectrum  except for the ground state of $H_1$. Suppose
that $\psi_1(x)$ is an eigenfunction of $H_1$ with eigenvalue
$E_1$. Then in view of \eq{f2} we have
    \be
    A^\dagger A \psi_1(x)=E_1\psi_1(x)\;.
    \label{f5.1}
    \ee
Using \eq{f4} it is easily seen that $A\psi_1(x)$ is an
eigenfunction of $H_2$ with eigenvalue $E_1$, provided that
$\psi_1(x)$ is not the ground state wave function.

One can repeat the above procedure for $H_2$ and get the
supersymmetric partner of $H_2$. Then one arrives at a hierarchy
of Hamiltonians which their spectrum are essentially the same.
Now suppose that the potentials $V_1(x)$ and $V_2(x)$ depend on
some parameters which we show them in a compact form by $a$. The
shape invariance condition is
    \be
    V_2(x,a_1)=V_1(x,a_2)+R(a_1)\;,
    \label{sic}
    \ee
where $a_2=F(a_1)$ is a function of $a_1$ and $R(a_1)$ is
independent of $x$. If these conditions are fulfilled, the
spectrum of $H_1$ can be found algebraically~\cite{rev}:
    \be
    E_n=\sum^n_{i=1}R(a_n)\;.
    \label{eev}
    \ee
In what follows, we use this method to find the exact solutions
of effective mass Schr\"{o}dinger equation considering the super
potentials in the ansatz:
    \be
    W(x,a)=ag(x)+f(x)+\frac{1}{a}h(x)\;.
    \label{ansatz}
    \ee
Inserting $W(x,a)$ from \eq{ansatz} in \eqs{f3.2} and \e{f5}, the
shape invariance condition \e{sic} gives
    \bea
    &&(a_2^2-a_1^2)g^2(x)+\left(\frac{1}{a_2^2}-\frac{1}{a_1^2}\right)h^2(x)+
    2(a_2-a_1)g(x)f(x)+2\left(\frac{1}{a_2}-\frac{1}{a_1}\right)h(x)f(x)\nonumber\\
    &&-\left[U(x)(a_2-a_1)g(x)+U(x)\left(\frac{1}{a_2}-\frac{1}{a_1}\right)h(x)\right]'
    -2U(x)\left[a_1g(x)+\frac{1}{a_1}h(x)+f(x)\right]'\nonumber\\
    &&+U(x)U^{''}(x)+R(a_1)=0\;.
    \label{sic2}
    \eea
Before starting  to study \eq{sic2}, we make a definition for the
future convenience:
    \be
    Y(x):=\int^x\frac{dx'}{U(x')}\;.
    \label{intu}
    \ee
There is also a comment in order on the square integrability of
the wave functions. According to \eq{f2} the ground state of
$H_1$ is given by $A\psi_0(x)=0$. This, together with definition
\e{f3} leads to
    \be
    \psi_0(x)\sim \exp \left[-\int^x
    \frac{W(x')}{U(x')}dx'\right]\;.
    \label{f31}
    \ee
Then the square integrability of $\psi_0(x)$ leads to
    \be
    \int^\infty_{-\infty}\psi_0^2(x) dx =\int^\infty_{-\infty}\exp
    \left[-2\int^x\frac{W(x')}{U(x')}dx'\right]<\infty\;.
    \label{sqint}
    \ee
This condition  puts some restrictions on acceptable $W(x)$ and
$U(x)$. In fact the following condition must be fulfilled
    \be
    \lim_{x\rightarrow \pm
    \infty}\int^x\frac{W(x')}{U(x')}dx'=+\infty\;.
    \label{sqint2}
    \ee
If $\frac{W(x)}{U(x)}$ has definite signs for  $x\rightarrow \pm
\infty$, then there is a simple condition which guarantees
\eq{sqint2}~\cite{susy1}. Let us define
    \be
    \left(\frac{W}{U}\right)_\pm:=\lim_{x\rightarrow \pm
    \infty}\frac{W(x)}{U(x)}\;.
    \label{cond}
    \ee
Then, to fulfill \eq{sqint2}, it is sufficient that
    \be
    {\rm
    sign}\left(\frac{W}{U}\right)_+=-{\rm
    sign}\left(\frac{W}{U}\right)_-\;.
    \label{cond2}
    \ee
It can be shown that this condition also ensures the square
integrability of excited states wave functions~\cite{rev}.
\section{GSIPs of type $W(x,a)=f(x)+a$}
In this section we consider GSIPs whose superpotentials are of
the form
    \be
    W(x,a)=f(x)+a\;.
    \label{s1.0}
    \ee
This means that in \eq{ansatz} we have put $g(x)=1$ and $h(x)=0$.
Then the shape invariance condition gives
    \be
    U(x)\left[U'(x)-2f(x)\right]'-(a_2-a_1)\left[U'(x)-2f(x)\right]+(a_2^2-a_1^2)+R(a_1)=0\;.
    \label{s1.1}
    \ee
In this equation $U(x)$ and $f(x)$ are independent of parameter
$a_1$. Therefore there are  only two ways for \eq{s1.1} to be
consistent. The first way is that $U'(x)-2f(x)=const.=:2u_0$. Then
the superpotential is given by $W(x,a)=\frac{1}{2}U'(x)+a-u_0$.
and \eq{s1.1} implies that $R(a_1)=(a_1^2-a_2^2)-2u_0(a_1-a_2)$.
Absorbing $u_0$ in $a$, one gets $R(a_1)=(a_1^2-a_2^2)$. Therefore
in this case $a_2$ is an arbitrary function of $a_1$ provided
$R(a_1)=a_1^2-a_2^2>0$ to ensure that the energy eigenvalues are
non negative.

The second way  is that $a_1-a_2$ and $(a_2^2-a_1^2)+R(a_1)$ be
independent of parameter, i.~e.
    \bea
    R(a_1)&=&\alpha(2a_1-\alpha)+R_0\;, \label{s1.2}\\
    a_2&=&a_1-\alpha\;, \label{s1.2.1}
    \eea
where $\alpha$ and $R_0$ are some real constants. Using these
conditions and defining $u(x):=U'(x)-2f(x)$ we can write \eq{s1.1}
in the following form
    \be
    U(x)u'(x)+\alpha u(x)+R_0=0\;.
    \label{s1.3}
    \ee
Before trying to solve this equation let us have a look at the
spectrum of $H_1$. The energy eigenvalues of $H_1$ are given by
    \be
    E_n=\sum^n_{i=1}R(a_i)=n(R_0+2a_1\alpha-n\alpha^2)\;.
    \label{en1}
    \ee
This equation shows that for $\alpha=0$ the energy levels are
equi-spaced. Due to  the importance of systems with equi-spaced
energy levels we consider two distinct cases corresponding to
$\alpha=0$ and $\alpha \ne 0$ separately. We call solutions with
equi-spaced energy levels {\it oscillator like solutions}.
\subsection{Oscillator like Solutions}
For $\alpha=0$ \eq{s1.3} gives
    \be
    u(x)=-R_0 Y(x)\;,
    \label{s11}
    \ee
where $Y(x)$ is defined by \eq{intu}. Therefore the superpotential
is
    \be
    W(x,a)=\frac{1}{2}\left[U'(x)+R_0 Y(x)\right]+a\;,
    \label{s1.8}
    \ee
and the potential $V_1(x,a_1)$ is given by
    \be
    V_1(x,a)=\frac{1}{4}\left[(R_0Y(x)+2a)^2-2R_0\right]+V_0(x)\;,
    \label{V11}
    \ee
in which
    \be
    V_0(x):=-\frac{1}{4}U'^2(x)-\frac{1}{2}U(x)U^{''}(x)\;.
    \label{deff}
    \ee
The energy eigenvalues  are $E_n=nR_0\;,n=0,1,2,\cdots $. Since
the energy eigenvalues are non negative we should have $R_0>0$.
because the energy eigenvalues are non negative.

Now we can obtain the zero energy eigenstate using \eq{f31}
    \be
    \psi_0(x)\sim |U(x)|^{-1/2}\exp\left(-\frac{R_0}{4}Y^2(x)-a
    Y(x)\right)\;.
    \label{s1.9}
    \ee
It can be  easily verified that the ground state wave function is
normalizable i.~e. \eq{sqint} is fulfilled. The results of this
subsection is summarized in table~\ref{tab1}.

\subsubsection{Examples}
\noindent {\it Example 1:} If $U(x)=\frac{1}{\sqrt {2m}}=const$,
the system is a  harmonic oscillator; the most simple system with
equi-spaced energy levels.

\noindent {\it Example 2:} For $U(x)=\frac{1}{2x}$ we have
$Y(x)=x^2$. The effective mass and potential are given by
    \bea
    &&m(x)=2x^2\;, \label{ex112}\\
    &&V_1(x,a)=\left(\frac{R_0}{2}x^2+a\right)^2-\frac{5}{16x^4}-\frac{R_0}{2}\;.
    \label{ex113}
    \eea
\noindent {\it Example 3:} For $U(x)=\frac{1}{\cosh x}$ we have
$Y(x)=\sinh x$ and $m(x)=\frac{1}{2}\cosh^2 x$. This form of
position dependent mass is considered by some authors in graded
alloys~\cite{ap3}. The potential is given by
    \be
    V_1(x,a)=\frac{1}{4}\left[(R_0\sinh x + 2a)^2+\frac{\sinh^2 x
    -2}{\cosh^4 x} -2R_0\right]\;.
    \label{ex114}
    \ee
\subsection{Exponential Solutions}

If $\alpha\ne 0$, by redefinition of the function $f(x)$ and the
parameter $a$ we can put $R_0=0$. The redefinition is given by
the following equations
    \bea
    a& \rightarrow &
    \tilde{a}=a+\frac{R_0}{2\alpha}\;,\label{redef1}\\
    f(x) & \rightarrow & \tilde{f(x)}=f(x)-\frac{R_0}{2\alpha}\;.\label{redef2}
    \eea
Under this redefinition, $W(x,a)$ does not change but $R(a_1)$
changes as
    \be
    R(a_1) \rightarrow
    \tilde{R}(\tilde{a_1})=\alpha(2\tilde{a_1}-\alpha)\;,
    \label{redef3}
    \ee
and \eq{s1.3} can be rewritten as
    \be
    U(x)\tilde{u}'(x)+\alpha \tilde{u}(x)=0\;.
    \label{s1.3t}
    \ee
We conclude that for $\alpha \ne 0$ one can always put $R_0=0$.
The solution of \eq{s1.3t} is
    \be
    \tilde{u}(x)=u_0\exp\left(-\alpha Y(x)\right)\;.
    \label{s1.5}
    \ee
We call these  solutions, {\it exponential solutions}. The
superpotential  $W(x,a)$ is therefore
    \be
    W(x,a)=\frac{1}{2}\left[U'(x)-u_0\exp\left(-\alpha
    Y(x)\right)\right]+a\;,
    \label{s1.6}
    \ee
and the potential is given by
    \be
    V_1(x,a)=\frac{1}{4}\left[(u_0e^{-\alpha Y(x)}-2a)^2-2\alpha u_0 e^{-\alpha
    Y(x)}\right]+V_0(x)\;,
    \label{vv}
    \ee
in which $V_0(x)$ is defined in \eq{deff}. The energy eigenvalues
are
    \be
    E_n=\alpha n(2a_1-\alpha n)\;.
    \label{en2}
    \ee
The  ground state which is given by \eq{f31} takes the following
form
    \be
    \psi_0(x)\sim |U(x)|^{-1/2}\exp\left(\frac{-u(x)}{2\alpha}-a_1
    Y(x)\right)\;.
    \label{s1.10}
    \ee
The above  results are summarized in the table~\ref{tab2}.

\subsubsection*{Examples}
\noindent {\it Example 1:} $U(x)=U_0=const$. With this choice for
$U(x)$ we arrive at  the Morse potential for ordinary
Schr{\"o}dinger equation which is a well known SIP~\cite{rev}.

\noindent {\it Example 2:} $U(x)=-\frac{\alpha}{2}x$. In this
case, choosing $u_0=1$ in \eq{s1.5}, one gets
    \bea
    &&W(x,a)=-\frac{1}{4}\left(\alpha+2x^2\right)+a\;,
    \label{s1.12}\\
    &&V_1(x,a)=\frac{1}{4}x^4-(\alpha/2+a)x^2
    -\frac{\alpha^2}{16}+a^2\;, \label{s1.13}\\
    &&\psi_0(x)\sim
    |x|^{\frac{2a_1}{\alpha}-\frac{1}{2}}\exp\left(-\frac{x^2}{2\alpha}\right)\;.
    \label{s1.14}
    \eea
\section{GSIPs of Type $W(x,a)=ag(x)+f(x)$}
This ansatz is obtained putting $h(x)=0$ in \eq{ansatz}. Applying
the shape invariance condition  we arrive at
    \bea
    &&(a_2^2-a_1^2)g^2(x)+(a_2-a_1)g(x)\left[2f(x)-U'(x)\right]\nn\\
    &&-(a_1+a_2)U(x)g'(x)
    -U(x)\left[2f(x)-U'(x)\right]'+R(a_1)=0\;.
    \label{s2.1}
    \eea
Assuming that $a_1-a_2=\alpha$, one gets
    \bea
    &&\alpha^2g^2(x)-2\alpha
    f(x)g(x)+\alpha\left[U(x)g(x)\right]'-2U(x)f'(x)+U(x)U^{''}(x)\nn\\
    &&-2a_1\left(\alpha g^2(x)+U(x)g'(x)\right)+R(a_1)=0\;.
    \label{s2.2}
    \eea
Since $U(x)$, $f(x)$, $g(x)$, and $\alpha$ are parameter
independent, we conclude from the above equation that
    \bea
    &&\alpha^2g^2(x)-2\alpha
    f(x)g(x)+\alpha\left[U(x)g(x)\right]'-2U(x)f'(x)+U(x)U^{''}(x)+C_1=0\;,
    \label{s2.3}\\
    &&-2a_1\left(\alpha g^2(x)+U(x)g'(x)\right)+R(a_1)-C_1=0\;.
    \label{s2.4}
    \eea
In the same way \eq{s2.4} implies that
    \bea
    &&\alpha g^2(x)+U(x)g'(x)+C_2=0\;, \label{s2.5}\\
    &&R(a_1)=C_1- 2 a_1 C_2\;.\label{s2.6}
    \eea
Therefore energy eigenvalues are
    \be
    E_n=n^2\alpha C_2+n[C_1-(2a_1+\alpha)C_2]\;.
    \label{s2.4.1}
    \ee
Using \eq{s2.5} in \eq{s2.3} one finds
    \be
    \alpha g(x)u(x)+U(x)u'(x)+C_1-\alpha C_2=0\;,
    \label{s2.7}
    \ee
where $u(x)=U'(x)-2f(x)$. \eq{s2.4.1} indicates that for $\alpha
C_2=0$ the energy eigenvalues are equi-spaced. In the following
we discuss three different families of solutions corresponding to
$\alpha C_2=0$, $\alpha C_2>0$, and $\alpha C_2<0$.
\subsection{Oscillator like solutions ($\alpha C_2=0$)}
These are solutions with equi-spaced  spectrum (see \eq{s2.4.1}).
Three different cases can be realized as follows,

{\it Case 1:} $\alpha=C_2=0$. In this case according to \eq{s2.5},
$g(x)$ should be a constant function of $x$. We have discussed
this case in section 3.

{\it Case 2:} $\alpha=0$ and $C_2\ne 0$. In this case \eqs{s2.5}
and \e{s2.7} are easily solved and we get $g(x)=-C_2Y(x)$ and
$u(x)=-C_1 Y(x)$. Therefore,
    \bea
    &&W(x,a)=\frac{1}{2}U'(x)+\frac{1}{2}(C_1-2aC_2)Y(x)\;,\label{add1}\\
    &&E_n=n((C_1-2a_1C_2)\;. \label{add2}
    \eea
One can redefine the parameter $a$ as $a\rightarrow
\tilde{a}=\frac{C_1}{2}-aC_2$. The results are summarized in the
table~\ref{tab3}.

{\it Case 3:} $\alpha \ne 0$ and $C_2=0$. In this case from
\eq{s2.5} one finds
    \be
    g(x)=\frac{1}{\alpha Y(x)}\;.
    \label{s2.8}
    \ee
Then substituting this solution in \eq{s2.7} we have
    \be
    u(x)=-\frac{C_1}{2}Y(x)-\frac{C_3}{Y(x)}\;,
    \label{s2.9}
    \ee
where $C_3$ is a constant. In fact $C_3$ is not so important and
can be absorbed in the parameter $a$. See table~\ref{tab4}
\subsection{Trigonometric Solutions  ($\alpha C_2>0$)}
Putting $k:={\sqrt {C_2\alpha}}$ it is straightforward to solve
\eqs{s2.5} and \e{s2.7} to get
    \be
    g(x)=-\frac{k}{\alpha}\tan \left[k Y(x)\right]\;,
    \label{s2.10}
    \ee
and
    \be
    u(x)=-\frac{C_1-\alpha C_2}{k} \tan\left[
    k Y(x)\right]-C_3 \sec\left[k Y(x)\right]\;.
    \label{s2.11}
    \ee
Then the superpotential is obtained:
    \be
    W(x,a)=(-\frac{k a}{\alpha}+\frac{C_1-k^2}{2k})\tan(k
    Y)+\frac{C_3}{2} \sec(k Y)\;.
    \label{add3}
    \ee
Redefining the parameter $a$ as
    \be
    a \rightarrow \tilde{a}=-\frac{k
    a}{\alpha}+\frac{C_1-k^2}{2k}\;,
    \label{add4}
    \ee
we have $\tilde{\alpha}:=\tilde{a_1}-\tilde{a_2}=-k$. Using this
redefinition one can write the general form of superpotential as
    \be
    W(x,a)=-a\tan(\alpha Y)+b\sec(\alpha Y)+\frac{1}{2}U'(x)\;,
    \label{s2.14}
    \ee
where $b=-\frac{C_3}{2}$. The energy eigenvalues are given by
    \be
    E_n=n\alpha(n\alpha-2a_1)\;.
    \label{s2.16}
    \ee
Finally the ground state is given by
    \be
    \psi_0(x)\sim |U(x)|^{\frac{-1}{2}}|\sec(\alpha
    Y)|^{\frac{a}{\alpha}}|\tan(\alpha Y)+\sec(\alpha
    Y)|^{-\frac{b}{\alpha}}\;.
    \label{s2.17}
    \ee

It is remarkable that for $U(x)=const.$ the above solution reduces
to the Scarf I potential which is a SIP for ordinary
Schr{\"o}dinger equatioin~\cite{rev}. The results of this
subsection is summarized in table~\ref{tab5}
\subsection{Hyperbolic Solutions ($\alpha C_2<0$)}
Considering the definition  $k:={\sqrt {-C_2\alpha}}$ one can
solve \eqs{s2.5} and \e{s2.7} to get
    \be
    g(x)=\frac{k}{\alpha}\tanh \left[k Y(x)\right]\;,
    \label{s2.12}
    \ee
and
    \be
    u(x)=-\frac{C_1- \alpha C_2}{k}\tanh\left[
    k Y(x)\right]-C_3 \sech\left[ k Y(x)\right]\;.
    \label{s2.13}
    \ee
Then the superpotential can be found as
    \be
    W(x,a)=(\frac{k a}{\alpha}+\frac{C_1+k^2}{2k})\tanh(k
    Y)+\frac{C_3}{2} \sech(k Y)\;.
    \label{ad3}
    \ee
Now we redefine the parameter $a$ as
    \be
    a \rightarrow \tilde{a}=\frac{k
    a}{\alpha}+\frac{C_1+k^2}{2k}\;.
    \label{ad4}
    \ee
After this redefinition of $a$ we have
$\tilde{\alpha}:=\tilde{a_1}-\tilde{a_2}=k$. Therefore the general
form of superpotential can be given as
    \be
    W(x,a)=a\tanh(\alpha Y)+b\sech(\alpha Y)+\frac{1}{2}U'(x)\;,
    \label{ad5}
    \ee
where $b=\frac{C_3}{2}$. The energy eigenvalues are given by
    \be
    E_n=n\alpha(n\alpha-2a_1)\;,
    \label{ad6}
    \ee
and  the ground state is
    \be
    \psi_0(x)\sim |U(x)|^{\frac{-1}{2}}|\sech(\alpha
    Y)|^{\frac{-a}{\alpha}}|\tanh(\alpha Y)+\sech(\alpha
    Y)|^{-\frac{b}{\alpha}}\;.
    \label{ad7}
    \ee

For $U(x)=const.$ the above solution reduces to the Scarf II
potential which is a SIP for ordinary Schr{\"o}dinger
equatioin~\cite{rev}. The results of this subsection are
summarized in table~\ref{tab6}

\section{Concluding Remarks}
In this paper we studied exact solutions of effective mass
Schr\"{o}dinger equation using the mathod of shape invariant
potentials. We considered an ansatz in which the superpotential
takes the form
 \be
 W(x,a)=ag(x)+f(x)+\frac{h(x)}{a}.
 \ee
 Assuming $h(x)$  be vanishing, the exact solutions for the Schr\"{o}dinger equation are
 studied and a wide class of solutions with equi-spaced spectrum are obtained.
 Following the general arguments of the paper, it can be verified
 that if $h(x)$ is non-vanishing then it should be simply a
 constant. In this case the shape invariance condition requires
 $m(x)$ to be constant. Thus if $h(x)\neq 0$, one obtains the SIPs
 for the ordinary Scr\"{o}dinger equation. The main results of
 the this study are summarized in tables 1 -- 6.

\section*{Acknowledgment}
Financial supports from Isfahan University of Technology (IUT) is
acknowledged.

\newpage
    \bt
    & \\
    Superpotential & $W(x,a)=\frac{1}{2}\left[U'(x)+R_0
    Y(x)\right]+a$ \\
    & \\ \hline
    &\\
    Potential &
    $V_1(x,a)=\frac{1}{4}\left[(R_0Y(x)+2a)^2-2R_0\right]+V_0(x)$\\
    &  \\ \hline
    & \\
    Energy eigenvalues & $E_n=nR_0$\\ &\\ \hline
    & \\
    Ground state &$\psi_0(x)=N|U(x)|^{-1/2}\exp\left(-\frac{R_0}{4}Y^2(x)-a
    Y(x)\right)$\\&  \\ \hline
    \et{GSIPs of type $W(x,a)=f(x)+a$; Oscillator like solutions}{tab1}
    \bt & \\
    Superpotential & $W(x,a)=\frac{1}{2}\left[U'(x)-u_0\exp\left(-\alpha
    Y(x)\right)\right]+a$\\&\\ \hline &\\
    Potential & $V_1(x,a)=\frac{1}{4}\left[(u_0e^{-\alpha Y(x)}
    -2a)^2-2\alpha u_0 e^{-\alpha
    Y(x)}
    \right]+V_0(x)$\\&\\ \hline & \\
    Energy eigenvalues & $E_n=\alpha n(2a_1-\alpha n)$\\ & \\
    \hline & \\
    Ground state & $\psi_0(x)\sim
    |U(x)|^{-1/2}e^{\frac{-u(x)}{2\alpha}u(x)-a_1
    Y(x)}$\\ & \\ \hline
    \et{GSIPs of type $W(x,a)=f(x)+a$; Exponential solutions}{tab2}
    \bt
    &\\
    Superpotential & $W(x,a)=\frac{1}{2}U'(x)+a Y(x)$\\&\\
    \hline &\\
    Potential & $V_1(x,a)=a^2 Y^2(x)-a+V_0(x)$\\&\\
    \hline &\\
    Spectrum & $E_n=2a_1 n$\\ &\\ \hline &\\
    Ground state & $\psi_0(x)\sim |U(x)|^{-1/2}\exp\left(\frac{-a}{2}Y^2(x)\right)$\\ &\\
    \hline
    \et{GSIPs of type $W(x,a)=ag(x)+f(x)$; Oscillator like solutions (I)}{tab3}
    \bt
    &\\ Superpotential & $W(x,a)=\frac{1}{2}U'(x)+\frac{C_1}{4}Y(x)+\frac{a}{\alpha Y(x)}$\\&\\
     \hline &\\
    Potential & $V_1(x,a)=\frac{C_1^2}{16} Y^2(x)
    +\left(\frac{a^2}{\alpha^2}+\frac{a}{\alpha}\right)\frac{1}{Y^2(x)}+
    \frac{C_1}{2}\left(\frac{a}{\alpha}-\frac{1}{2}\right)+V_0(x)$\\& \\
    \hline &\\
    Spectrum & $E_n=C_1 n$\\&\\ \hline &\\
    Ground state & $\psi_0(x)\sim |U(x)|^{-1/2}|Y(x)|^{-\frac{a}{\alpha}}\exp(-\frac{C_1}{8}
    Y^2(x))$\\&\\ \hline
    \et{GSIPs of type $W(x,a)=ag(x)+f(x)$; Oscillator like solutions (II)}{tab4}
    \bt
    &\\ Superpotential & $W(x,a)=-a\tan(\alpha Y)+b\sec(\alpha Y)+\frac{1}{2}U'(x)$\\&\\
     \hline &\\
    Potential & $V_1(x,a)=(a\tan(\alpha Y)-b\sec(\alpha Y))^2$\\ & $-\alpha\sec(\alpha Y)
    (b\tan(\alpha Y)-a\sec(\alpha Y))+V_0(x)$\\& \\
    \hline &\\
    Spectrum & $E_n=n\alpha(n\alpha-2a_1)$\\&\\ \hline &\\
    Ground state & $\psi_0(x)\sim |U(x)|^{\frac{-1}{2}}|\sec(\alpha
    Y)|^{\frac{a}{\alpha}}|\tan(\alpha Y)+\sec(\alpha
    Y)|^{-\frac{b}{\alpha}}$\\&\\ \hline
    \et{GSIPs of type $W(x,a)=ag(x)+f(x)$; Trigonometric solutions}{tab5}
    \bt
    &\\ Superpotential & $W(x,a)=a\tanh(\alpha Y)+b\sech(\alpha Y)+\frac{1}{2}U'(x)$\\&\\
     \hline &\\
    Potential & $V_1(x,a)=(a\tanh(\alpha Y)-b\sech(\alpha Y))^2$\\&
    $+\alpha\sech(\alpha Y)
    (b\tanh(\alpha Y)-a\sech(\alpha Y))+V_0(x)$\\& \\
    \hline &\\
    Spectrum & $E_n=n\alpha(n\alpha-2a_1)$\\&\\ \hline &\\
    Ground state & $\psi_0(x)\sim |U(x)|^{\frac{-1}{2}}|\sech(\alpha
    Y)|^{-\frac{a}{\alpha}}|\tanh(\alpha Y)+\sech(\alpha
    Y)|^{-\frac{b}{\alpha}}$\\&\\ \hline
    \et{GSIPs of type $W(x,a)=ag(x)+f(x)$; Hyperbolic solutions}{tab6}

\newpage
\newpage


\newpage
\begin{thebibliography}{99}
\bibitem{1}E. Schr{\"o}dinger, Proc. Roy. Irish Acad.~{\bf A46}, 9 (1940).
\bibitem{factorization}L. Infeld and T. E. Hull, Rev. Mod. Phys.
{\bf 23}, 21 (1951).
\bibitem{witten}E. Witten, Nucl. Phys. {\bf B188}, 513 (1981).
\bibitem{sip}L. Gendenshtein, JETP Lett. {\bf 38}, 356 (1983).
\bibitem{rev}f. Cooper, A. Khare, and U. Sukhatme, Phys. Rep. {\bf
251}, 267 (1995).
\bibitem{fp} M. Bernstein and L. S. Brown, Phys. Rev. Lett. {\bf
52}, 1933 (1984).
\bibitem{bd}D. J. BenDaniel and C. B. Duke, Phys. Rev. {\bf
152},683 (1966).
\bibitem{ap1} G. Bastard, {\it Wave Mechanics Applied to
Semiconductor Heterostructures}, Les Editions de Physique, Les
ulis, France (1988).
\bibitem{ap2} D. Bessis and G. A. Mezincescu, Microelectronic J.
{\bf 30}, 953 (1999).
\bibitem{ap3}V. Milanovi\'{c} and Z. Ikoni\'{c}, Phys.
Rev. {\bf B54}, 1998 (1996).
\bibitem{ex1} B. Roy and P. Roy, J. Phys. A: Math. Gen. {\bf
35}, 3961 (2002).
\bibitem{ex2} B\"{u}lent G\"{o}n\"{u}l, Okan \"{O}zer,
Be\c{s}ire G\"{o}n\"{u}l, and Fatma \"{U}zg\"{u}n, Mod. Phys.
Lett.  {\bf A17}, 2453 (2002).
\bibitem{ex3} A. D. Alhaidari, Phys. Rev. {\bf A66}, 042116
(2002).
\bibitem{susy1}L. E. Gendenshtein and I. V. Krive, Sov. Phys. Ups.
{\bf 28}, 645 (1985).
\end{thebibliography}
\end{document}